# Noncentralized Cryptocurrency wtih No Blockchain

by Qian Xiaochao

fxxfzx@qq.com

btc: 18Bf718HZNboFvEqQuhJCTdKJtM8YxHPZW

Sept. 2016, final

**Abstract.** We give an explicit definition of decentralization and show you that decentralization is almost impossible for the current stage and Bitcoin is the first truly noncentralized currency in the currency history. We propose a new framework of noncentralized cryptocurrency system with an assumption of the existence of a weak adversary for a bank alliance. It abandons the mining process and blockchain, and removes history transactions from data synchronization. We propose a consensus algorithm named “Converged Consensus” for a noncentralized cryptocurrency system.

**Keywords:** Cryptocurrency,Converged Consensus, Bitcoin, Noncentralization

## Introduction

Bitcoin is a peer to peer distributed digital currency system whose implementation is mainly based on cryptography. As a decentralized cryptocurrency, it has attracted lots of attention and has been widely adopted over the whole world. It is proposed by Satoshi Nakamoto in 2008 and created in 2009. Its goal is to build a practical decentralized cryptocurrency system. To achieve this goal, it adopts a proof of work mechanism to distribute the coins and validate the transactions. This mechanism is implemented by a sophisticated design of block and blockchain. The system adopts a longest blockchain principle to represent the consensus of users about the general ledger.

In the Bitcoin system every transaction is stored in the blockchain. Over time, the size of the blockchain grows very fast. Ordinary users hate to verify transactions and synchronize the whole blockchain and the trading between Bitcoin and some fiat currency is invietable, so that they have to rely on some exchange hub which leads to additional safety dependence which severely compromises the safety of Bitcoins. The Mt.Gox event broke out against such a background. Many innocent users lost their Bitcoins. In addition, the intensive mining process results in a colossal waste of electricity and computing resources.

The proof of work mechanism makes the Bitcoin system expose itself to any potential superiority computing power attacker who may even hold no Bitcoin. There is a bad news about the Bitcoin system in Jun 2014, a major mining pool ghash.io had ever controlled more than 50% of computing power. This is a dangerous sign and is unacceptable for an alleged decentralized cryptocurrency system. Various of phenomenon express that the feasibility of a decentralized cryptocurrency is questionable. The main content of this paper is divided to two parts. We first try to show you that decentralization is almost impossible for the current stage in Part I and we propose a new framework of noncentralized cryptocurrency system for a bank alliance in Part II.

## Noncentralization

**Definition Centralization**: A cryptocurrency system which is totally controlled by one single entity.

**Definition Decentralization**: A cryptocurrency system which has only one safety dependency that is the private key.

**Definition Noncentralization**: A cryptocurrency system which is neither centralized nor decentralized.

Simply speaking, Decentralization means no trust; Noncentralization means partial trust and Centralization means fully trust.

## Mining Centralization

The longest chain principle of Bitcoin provides a winner-take-all competition mechanism. For any miner, the best strategy is to try its best to get stronger or to merge with other miners to get itself stronger. As long as there are more than one miner (a mining pool is deemed as a miner) such fierce competition won't cease. Thus the mining system has strong incentive to get centralized. The equilibrium result is that there is only one miner left.

Although there are still more than one major mining pools at present, the weak pools are going to disappear because they are not able to get enough rewards to cover the cost for running a mining pool. The number of mining pools is going down. Even though the number won't reduce to one, how could we know the major mining pools are not controlled by some single underground big boss? This is called Sybil attack which means mining pools with different names are merely sybils of some underground big boss.

According to previous definition of Decentralization and Noncentralization, Bitcoin is obviously not decentralized as some fans claimed but running as a noncentralized system.

As can be seen, essentially, Satoshi's competitive blockchain solution to decentralization is centralization which is a trivial solution. As a decentralized cryptocurrency, the Bitcoin system is clearly failed, however, a failed experiment can provide us much valuable information.

## First Noncentralized Currency

Although mining centralization is inevitable, Bitcoin is still a truly noncentralized cryptocurrency system. In fact, the mining center will control most part of the Bitcoin system but not the whole system. In a typical centralized currency system, the central bank has super power of transferring money from any account to any other account, while the mining center doesn't has such super power and it's possible for stakeholders to fork the Bitcoin system whenever they find the mining center is no longer trustworthy. So Bitcoin is not totally controlled by the mining center. According to our previous definition, Bitcoin system is a noncentralized cryptocurrency and it is the first truly noncentralized currency in the currency history.

## Virtual General Adversary (VGA)

For any decentralized system we assume there is one single Virtual General Adversary which is trying to attack the system. "Virtual General" means it includes any factor (including communication problem and some abstract attack behaviors) which may lead to the failure of decentralization. Virtually, the adversary can do whatever it can do to attack the system. We consider 4 major attack behaviors.

①maliciously construct messages,
②maliciously schedule messages,
③sybil attack
④centralization

The Bitcoin system manages to build up strong resistance to ①, because constructing a block with high difficulty value is very expensive, nevertheless, it has very weak resistance to ②(e.g. a mining pool may withhold blocks it discovered) ③(we don't know whether there is a underground boss)④(there is strong incentive to get centralized). These mean that the blockchain mechanism merely partially meets the challenge ①.

To overcome any one of these 4 attack behaviors is a big challenge, let alone to overcome all of them simultaneously to build a truly decentralized cryptocurrency. Decentralization is too good to be true for the current stage. Probably, it belongs to the far future say the $22^{nd}$ century, by when some truly revolutionary breakthrough of communication technique or computational theory may have occurred.

To take the second best, we try to propose a new framework of noncentralized cryptocurrency system with an assumption of the existence of a weak adversary, which may be practical for the current stage.

When we talk about an adversary we consider an adversary quadruple Q:= <R,S,I,C> where R is the basic Resources (e.g. computing power, money) the adversary possesses, S is the ability the adversary Scheduling messages, I is the ability the adversary gathering Information of the system, C is the attraction degree of the system getting Centralized.

It’s hard to accurately quantify the value of Q and if we simply assume a super high value of Q then we always get negative result. Hence, we assume a weak virtual general adversary with a low Q value for our new system. How weak exactly? Honestly, we have no idea. Just like the Bitcoin system we need to do experiments to gather information and finetune the parameters.

We suppose the system is found by a bank alliance which includes N member banks all over the whole world. The banks are highly trusted, not completely trusted though. That accords with the assumption of a weak adversary. We name the new system X-Coin. Every expected event in the system has a timeout value, some of which are set globally the others are set locally.

**Balanceview**

Since we don't need mining anymore, we replace the concept of block and blockchain with normal data list.

**Definition** ***Balanceview*** *is a data list compounding a balance record list and a bank list.*

**Definition** ***Baseview*** *is a Balanceview based on which some important actions are taken.*

```
balance record
[
   pubkey,
   balance
]

balanceview
[
    bank list
    height,  //sequence number of a Balanceview increasing continuously from 0
    baseview hash, // refers to the Baseview of this Balanceview
    package hash,      // refers to the package which is actually calculated
    [record1, record2, record3],
]

transaction
[
    height,
    baseview hash,
    unique code,
    sender pubkey,
    receiver pubkey,
    volume,
    transaction fee,
    sender signature
```

]

transaction list
[
    [tx1, tx2, tx3... ],
    bank signature
]

An end user sets a max transaction fee he will pay, for every new transaction and sends it to a selected bank. The bank will pick up the transaction if it provides enough transaction fee. Every bank verifies and collects transactions and makes a transaction list and broadcasts the list to other banks.

**Definition agent** An agent is a bank which chooses to collect transaction lists from banks and make a transaction package and broadcasts the package to be verified and granted by banks.

A bank grants a package which shares the same Baseview by signing the package and sends a granter item back to the agent. Every bank freely sets a granting rule to select out packages which it wants to grant.

**Definition Package-51** An agent collects granter items from other banks. Once it collects over 51% of grants, it combines the transactions package and a granter items list to form a pakcage-51 and broadcasts this package-51 to announce a new candidate Balanceview. A new Balanceview can be calculated according to a package-51 and the last Baseview. The total transaction fees are divided by the ratio: p for the agent, q for the deputy bank, r equally divided to all banks.

granter item
[
    granter's pubkey,
    granter's signature   // signs a package
]

granter item list
[
    [item1, item2, item3... ],
]

transaction package
[
    height,
    baseview hash,
    pubkey of source node,
    transaction list,

signature,
]

package-51
[
package,
granter item list,
signature of agent
]

**Consensus**

It's possible that more than one package based on one Baseview get 51% of grants, consequently, more than one package-51 is broadcasted. Then the consensus of the new Balanceview can be split. This circumstance is analogous to a blockchain forks.

According to the longest chain principle, the Bitcoin system guarentee the termination of a consensus process, which means whenever you see a longer valid blockchain you decide on that blockchain. But the consistency is not 100% gaurenteed. Because multiple valid blocks at one height is possible and the virtual general adversary may mailously schedule the spreading of blocks. To prevent such uncertainty, a checkpoint mechanism and a confirmation delay of 6 blocks is introduced. Only decentralization guarantees both termination and consistency. Bitcoin is obviously not decentralized.

Taking the consensus standard of Bitcoin as a reference, X-Coin guarantees the termination and consistency with high probability. Our main idea of consensus algorithm is roughly as simple as following.

Suppose there are 100 red balls and 100 black balls in a bag, one randomly selects 5 balls, if red balls is more than black balls then one balck ball will turn to red ball, vise versa. He does this time and time again, eventually colors of all 200 balls will converge to one single color, ether all black or all red.

Suppose there are 100 balls in a bag, if one consecutively 20 times randomly selects 1 ball are all red balls. Then the probability of most of 100 balls are red balls is extremly high.

In practice, we present a Balanceview Indirectly denoted by ibv to save the space. When necessary, we reconstruct the Balanceview according to an ibv.

indirect presented Balanceview
[
height,
hash of package-51, // refers to a package-51 which is actually calculated
]

When a bank receives the first valid package-51 at its current height, it makes an ibv of this new Balanceview as its current ibv.

**Converged Consensus Algorithm:**

```
while no consecutive identical k ibv_update observed
     randomly sample n ibv // timeout gives a NULL ibv
     ibv_update = the most frequent ibv
     if ibv_update.height < current height
         discard   //we should never go backward
decide on ibv_update
if ibv_update == NULL or Baseview collision detected then alert
```

**Termination**: Every honest node eventually decides, with probability 1.
Since the adversary has limited ability of scheduling the messages, termination will be reached, sooner or later.

A bank reconstructs Baseview according to ibv_update. A bank's current Balance-view is volatile as the bank keeps updating. A Baseview is rather stable but still not permanent because the algorithm doesn't guarantee completely consistent.

Note that the distribution of ibv gets more and more consistent as the sampling and updating go and the speed of the converging process gets faster and faster due to the positive feedback effect which is a good merit for practical application.

To avoid double spending, referring to Bitcoin's strategy, we may delay the confirmation of transactions by several Baseviews.

A Baseview collision means some bank decides on different ibv at one height. If a bank detects a Baseview collision (theoretically possible, but very rare) at one height or it decides on a NULL ibv then a human intervention is needed.

**Meta data & meta-transaction**

We can adjust the major parameters of X-Coin system by a meta-transaction which requires 75% of grants to take effect.

We can adjust the ratio <p=0%, q=70%, r=30%> for dividing transaction fees which controls the intensity of competition among banks. We shouldn't give strong incentive to banks choosing to be an agent to increase the initial consistency as much as possible. So we set p=0% or a very small positive number. The ratio q controls the competition intensity among the banks serving end users. The ratio r provides the basic support to all banks to make the public contribution such as verification, granting and keeping sampling.

According to the Moore's Law, the cost of making public contribution will be trivial for a commercial bank and the banks are no longer a completely selfish entity hidden in the dark corner of the internet which can only be handled by Decentralization. We are now Noncentralization. The banks are legal public companies. They care about their public reputation and image. X-Coin system has weak resistance to a small number of dishonest banks.

We can adjust the sampling and termination parameters according to conditions of the environment. We can even inflate the money supply by a meta-transaction. If necessary, we can switch the system to a leader election mode, in which, we specifies a fixed bank as the agent to increase the consistency and efficiency or to fix some emergent problems.

We can make a statistic tool as a monitor model for a standard client for banks to monitor each other' behavior and a bank's extremely abnormal behaviors will be reported. We can use meta-transaction to remove a highly suspicious bank from the bank list as long as 75% of banks are convinced. Banks have sufficient incentive and social pressure to behaving normally and have little incentive to behave badly.

## Conclusion

We agree that maintaining a trustable financial system is very expensive. Satoshi's original idea is to trade expensive electricity and computing resources for Dencentralization. If the strategy does work then it worth the cost. Unfortunately, we have shown you that Bitcoin is actually running as a Noncentralized system and has obvious trend to evolve to a centralized system. So it's unnecessary to waste that much electricity and computing resources to do meaningless hashing.

As a decentralized system Bitcoin is obviously failed. For Decentralization the blockchain mechanism doesn't work and for Noncentralization the blockchain mechanism is unnecessary. The blockchain mechanism has been proven by the reality to be a trivial and wasteful design. We assert the blockchain mechanism will be a flash in the pan. The most valuable information we get from the failed Bitcoin experiment is that Bitcoin has shown us that a noncentralized cryptocurrency system is possible.

We have shown you that Decentralization is almost impossible in the current stage. Researchers should not claim a Decentralization and we should not believe in a claimed one. It's all Noncentralization. We have one question for those who claim a Decentralization. Whether the system has only one safety dependency? If the answer is no, what are other unknown unclear undefined safety dependencies? How could they ensure the safety without even knowing all safety dependencies? Actually, they couldn't give an accurate quantitified definition of the adversary. They don't know what their adversary exactly is and how strong their adversary exactly is. So, their calculations on the safety analysis are very questionable. Decentralization is simply too good to be true. It's not the time for Decentralization. We suggest give the noncentralized X-Coin system based on Converged Consensus a try.

Although it is extremely hard to build a truly decentralized cryptocurrency system, we human beings do have gotten one truly decentralized currency system. It is the gold system which has only one safety dependency that is YOU. Your gold is safe as long as you keep your gold well so the only safety dependency is YOU—the owner of gold. We think of the gold system as a NATUAL Decentralized currency system. The gold system itself is a special Balanceview which represents the NATUAL consensus of the distribution of fortune of all mankind.

Finally, we have a conjecture that the timeline of currency of mankind should be like this:

Natural Decentralization (gold) →

Artificial Centralization (fiat currency) →

*Artificial Noncentralization* (Bitcoin, X-Coin) → Artificial Decentralization (X-Gold, our ultimate goal)

## Appendix

Solution to the ball problem:
The changing of the state of all balls forms a Markov chain. Obviously, it's an absorbing Markov chain. Suppose its transition matrix P have t transient states (two colors) and r absorbing states (single color). Then:

$$P = \begin{pmatrix} Q & R \\ 0 & I_r \end{pmatrix}$$

where Q is a t-by-t matrix, R is a nonzero t-by-r matrix, 0 is an r-by-t zero matrix, and Ir is the r-by-r identity matrix. Thus, Q describes the probability of transitioning from some transient state to another while R describes the probability of transitioning from some transient state to some absorbing state.
n steps transition matrix:

$$P^n = \begin{pmatrix} Q & R \\ 0 & I_r \end{pmatrix}^n = \begin{pmatrix} Q^n & R' \\ 0 & I_r \end{pmatrix}$$

The probability that the process will be absorbed is 1. Because $\lim_{n\to\infty} Q^n = 0$